\documentstyle[preprint,aps,psfig]{revtex}  

\tighten       

\begin{document}

\draft
\preprint{Version of \today }

\title{Time-periodic phases in populations of nonlinearly coupled
oscillators with bimodal frequency distributions}

\author{L.~L.\ Bonilla$^1$ \cite{bonilla:email}, C.~J.\ P\'erez Vicente$^2$ 
\cite{cjpv:email}, and
R.\ Spigler$^3$ \cite{spigler:email} }

\address{$^1$Universidad Carlos III de Madrid, Escuela Polit\'{e}cnica 
Superior, Butarque 15, 28911 Legan\'{e}s, Spain \\
$^2$Department de F\'{\i}sica Fonamental, Facultat de F\'{\i}sica, 
	Universitat de Barcelona, Diagonal 647, 08028 Barcelona, Spain\\
$^3$Dipartimento di Matematica, Universit\`a di 
``Roma Tre'', Via Corrado Segre 60, 00146 Roma, Italy
}







\maketitle

\begin{abstract}
The mean field Kuramoto model describing the synchronization of a 
population of phase oscillators with a bimodal frequency distribution 
is analyzed (by the method of multiple scales) near regions in its phase 
diagram corresponding to synchronization to phases with a time periodic 
order parameter. The richest behavior is found near the tricritical 
point were the incoherent, stationarily synchronized, ``traveling wave'' 
and ``standing wave'' phases coexist. The behavior near the tricritical
point can be extrapolated to the rest of the phase diagram. 
Direct Brownian simulation of the model confirms our findings.
\end{abstract}


\section{Introduction}

In recent years mathematical modeling and analysis of synchronization
phenomena received increased attention because of its occurrence in quite
different fields, such as solid state physics \cite{TS1,TS2,CDW},
biological systems \cite{WIN,STRO,MISTRO,GRAY}, chemical reactions
\cite{KURAM2}, etc. These phenomena can be modeled in terms of populations
of interacting, nonlinearly coupled oscillators as first proposed by
Winfree \cite{WIN}. While the dynamic behavior of a small number of
oscillators can be quite rich \cite{ARO}, here we are concerned 
with synchronization as a collective phenomenon for large populations 
of interacting oscillators \cite{STRO}. 

A simple model put forth by Kuramoto and Sakaguchi \cite{KURAM,SAK} (see
also \cite{STRO}), consists of a population of coupled phase oscillators,
$\theta_{i}(t)$, having natural frequencies $\omega_{i}$ distributed with
a given probability density $g(\omega)$

\begin{equation}
  \dot{\theta_{i}} = \omega_{i} + \xi_{i}(t) + \sum_{j=1}^{N}
	  K_{ij} \sin(\theta_{j} -\theta_{i}),	\quad\quad
i=1,\ldots,N.
\label{1} 
\end{equation}
Here $\xi_{i}$ are independent white noise processes with expected values
\begin{equation}
	 \langle \xi_{i}(t) \rangle = 0,    \ \ \
      \langle \xi_{i}(t)\xi_{j}(t') \rangle = 2 D \delta(t -t')\,
			   \delta_{ij}.\label{2}
\end{equation}
Thus each oscillator tries to run independently at its own frequency while
the coupling tends to synchronize it to all the others. When the coupling is
sufficiently weak the oscillators run incoherently whereas beyond a
certain threshold collective synchronization appears spontaneously.
So far, several particular prescriptions for the matrix $K_{ij}$ have been
considered. For instance, $K_{ij}=K>0$ only when $|i-j|=1$, and $K_{ij}=0$
otherwise (next-neighbor coupling) \cite{STROMI2}; $K_{ij}=K/N>0$
(mean-field coupling) \cite{KURAM,KURAM2}; hierarchical coupling
\cite{LUMER}; random long-range coupling \cite{BON2,PAB,BON3} or even
state dependent interactions \cite{SOMP}. In the mean-field case, the
model (\ref{1})-(\ref{2}) can be written in a convenient form, defining
the (complex-valued) order-parameter 
\begin{equation}
	    r e^{i\psi}=\frac{1}{N}\sum_{j=1}^{N} e^{i\theta_{j}},
   \label{3}
\end{equation}
where $r(t)\geq 0$ measures the phase coherence of the oscillators, and
$\psi(t)$ measures the average phase. Then eq. (\ref{1}) reads
\begin{equation}
    \dot{\theta_{i}}=\omega_{i} + Kr\sin(\psi - \theta_{i}) +
		    \xi_{i}(t),  \ \ \	i=1, 2, \ldots, N.
  \label{4}
\end{equation}
In the limit of infinitely many oscillators, $N \rightarrow \infty$, a
{\it nonlinear integro-differential} equation of the {\it Fokker-Planck}
type was derived \cite{BON1,STROMI} for the one-oscillator probability
density, $\rho(\theta,t,\omega)$,  
\begin{equation}
       \frac{\partial \rho}{\partial t}=D \frac{\partial^{2} \rho}{\partial 
\theta^{2}} -		\frac{\partial}{\partial \theta}(v \rho),\label{5}
\end{equation}
the drift-term being given by
\begin{equation}
	      v(\theta,t,\omega)=\omega + K r sin(\psi - \theta),\label{6}
\end{equation}
and the order-parameter amplitude by
\begin{equation}
    r e^{i \psi}= \int_{0}^{2 \pi}\!\!\int_{-\infty}^{+\infty}
	e^{i\theta} \rho(\theta,t,\omega) g(\omega)\,d\theta\,d\omega
.\label{7} 
\end{equation}
The probability density is required to be $2 \pi$-periodic as a function
of $\theta$ and normalized according to
\begin{equation}
       \int_{0}^{2 \pi} \rho(\theta,t,\omega) d\theta = 1.\label{8}
\end{equation}

   Mean-field models such as those described above were studied, e.g, by
Strogatz and Mirollo \cite{STROMI} in case the frequency distribution,
$g(\omega)$, has reflection symmetry, $g(-\omega) = g(\omega)$
and it is {\it unimodal} [$g(\omega)$ is non-increasing for $\omega>0$]. 
In \cite{STROMI}, the authors showed that for $K$ smaller than a certain
value $K_c$, the incoherent
equiprobability distribution, $\rho_0 \equiv 1/(2\pi)$, is {\it linearly
stable}, and linearly unstable for $K>K_c$. As $D\to 0+$, the incoherence
solution is still unstable for $K>K_c$ [$=2/\pi g(0)$ at $D=0$], but it
is neutrally stable for $K<K_c$: the whole spectrum of the equation 
linearized about $\rho_0$ collapses to the imaginary axis. In \cite{BNS}, 
the {\it nonlinear} stability issue was addressed, and the case of a {\it 
bimodal} frequency distribution was considered [$g(\omega)$ is even and it 
has maxima at $\omega = \pm \omega_0$]. In this case, new bifurcations appear, 
and bifurcating synchronized states have been asymptotically constructed in 
the neighborhood of the bifurcation values of the coupling strength. The
{\it nonlinear} stability properties of such solutions were also studied for 
the explicit discrete example $g(\omega) = {1\over 2}[\delta(\omega-\omega_0) 
+ \delta(\omega+\omega_0)]$, cf.~\cite{BNS}. A complete bifurcation study 
taking into account the symmetry properties of $g(\omega)$ was carried out
by Crawford, \cite{CRAW}. Similar results were obtained by Okuda and
Kuramoto in the related case of mutual entrainment between 
populations of coupled oscillators with different frequencies \cite{OKUDA}.
The main results concerning linear stability of incoherence with a 
bimodal discrete frequency distribution are summarized in Fig.1 
(cf. Fig.1, p.319 in \cite{BNS}). Also, in Fig.5, p.~327 of \cite{BNS} 
a {\it global} bifurcation diagram left unresolved the full behavior
of the oscillatory branch starting at K=4D.


The purpose of this paper is to complete the investigation started in
\cite{BNS}, analyzing in detail (asymptotically) the solution living
in the neighborhood of the {\it tricritical} point $(K/D=4,
 \omega_{0}/D=1)$ in the parameter space $(K/D, \omega_{0}/D)$, Fig.1. It
turns out that such a task is far from being merely a detail, since
technical difficulties are nontrivial at all, and results
 allow to complete the conjectured diagram in Fig.4 as shown in Fig.5 below.
In Section II, a {\it two-time} analysis for the Hopf bifurcation, already
developed
in \cite{BNS}, is revisited; in Section III, a {\it multiscale} analysis is
performed near the
tricritical point, generalizing the asymptotic analysis earlier accomplished in
\cite{BNS}. The corresponding bifurcation equations have been solved
recasting the problem into a general formalism due to Dangelmayr and
Knobloch~\cite{DK}. Numerical results designed to confirm the previous 
findings are presented in Section IV, and these are summarized along with 
the analytical results in Section V. 

\section{Two-time scale analysis for the Hopf bifurcation}
\subsection{
Linearized problems}
Here we revisit certain results given in \cite{BNS}. In the Hopf analysis
conducted there, degeneracy of an eigenvalue of multiplicity two was
overlooked, as pointed out by Crawford \cite{CRAW}. We will recall here
the relevant points of the linear and nonlinear stability analysis near
the line $K=4D$ in Fig.~1 where a Hopf bifurcation from incoherence arises
for an even discrete bimodal frequency distribution $g(\omega)$. 
The linearized eigenvalue problem for this case may be obtained by inserting 
$\rho = 1/(2\pi) + \exp [\lambda t]\, \mu(\theta,\omega)$ in (\ref{5}) and 
(\ref{6}), and then ignoring terms nonlinear in $\mu$: 
\begin{eqnarray}
D \frac{\partial^{2} \mu}{\partial\theta^{2}} - \omega \frac{\partial 
\mu}{\partial \theta}  + {K\over 2\pi}\,		\mbox{Re}\, e^{-i\theta} 
\int_{0}^{2\pi} \int_{-\infty}^{+\infty} e^{i\theta^{\prime}}\,
\mu(\theta^{\prime},\omega^{\prime})\, g(\omega^{\prime}) d\theta^{\prime} 
d\omega^{\prime} = \lambda \mu, \label{linearized}\\
\int_0^{2\pi} \mu(\theta,\omega)\, d\theta = 0. \label{mu}
\end{eqnarray}
It can be shown that there are two eigenvalues $\lambda$, which solve the 
equation \cite{STROMI}:
\begin{equation}
{K\over 2}\,\int_{-\infty}^{+\infty} {g(\nu)\over \lambda + D + i\nu}\, 
d\nu = 1.\label{autov}
\end{equation}
They are explicitly given by~\cite{BNS}
\begin{equation}
       \lambda_{\pm} = -D + \frac{K}{4} \pm \frac{1}{4} \sqrt{K^2 - 16
		   \omega_{0}^2}\, ,\label{eigenv}
\end{equation}
when 
\begin{equation}
       g(\omega) = {1\over 2}\,[\delta(\omega - \omega_{0}) +\delta(\omega + 
\omega_{0})]\, . \label{g(w)}
\end{equation}
Fig.~1 is straightforwardly constructed from (\ref{eigenv}). Above the
dashed line, $4\omega_{0}> K$, and the eigenvalues are complex. Each
complex eigenvalue is doubly degenerate due to the reflection symmetry
of $g(\omega)$ \cite{CRAW}. By direct substitution into (\ref{linearized}), 
it can be checked that 
\begin{equation}
     \mu_{1} = \frac{e^{i\theta}}{D + \lambda + i\omega}\,,\quad \quad
\mu_2 =  \frac{e^{-i\theta}}{D + \lambda - i\omega},\label{eigenf}
\end{equation}
are two linearly independent eigenfunctions corresponding to the same semisimple
complex eigenvalue $\lambda$~\cite{CRAW}. They are related by the reflection
symmetry $\omega\to - \omega$, $\theta\to - \theta$. When $\lambda$ is real, 
these eigenfunctions are complex conjugate of each other. The eigenvalue 
$\lambda$ is no longer semisimple but it still has multiplicity two~\cite{CRAW}. 

\subsection{Two-time scale analysis}
Let us now recall how to use the method of multiple scales to 
construct the solution branches which bifurcate from incoherence at $K=4D$,
$\omega_0 >D$~\cite{BNS}. We define a small positive parameter $\varepsilon$ 
which measures the departure from the critical value $K_c =4D$ by 
\begin{equation}
K = K_c + \varepsilon^2 K_2, \quad\quad 0<\varepsilon\ll 1.\label{epsilon}
\end{equation}
$K_2 = \pm 1$ has to be determined later according to the direction of the
bifurcating branch and the scaling (\ref{epsilon}) will be justified later.
The probability density $\rho(\theta,t,\omega;\epsilon)$ will be sought 
for according to the Ansatz~\cite{BNS}:
 \begin{eqnarray}
      \rho (\theta, t; \omega; \varepsilon)
	 = \frac{1}{2 \pi} \exp \left\{ \sum_{j = 1}^{3} \varepsilon^j
\sigma_{j} (\theta,t,\tau) + O(\varepsilon^4) \right\}\, \label{expAnsatz}\\
\tau = (K-K_c)t = \varepsilon^2 K_2 t.\label{tau}
\end{eqnarray}
The rationale behind (\ref{expAnsatz}) is as follows. First of 
all, near $K=K_c$, small disturbances from incoherence decay or grow 
according to the values of the factor
\begin{eqnarray}
\exp [\lambda(K)t] \sim \exp \left[\mbox{Re} {\partial\lambda (K_{c})\over\partial
K}(K-K_c)t + i\, \mbox{Im}\lambda(K_{c})t\right]\, .\label{expfactor}
\end{eqnarray}
Here $\lambda(K)$ is given by (\ref{eigenv}) with $K$ given by (\ref{epsilon}) 
and $\omega_0>D$. Hence $\lambda(K) \sim \pm i \Omega +
\varepsilon^2 K_2 (1 \mp i D/\Omega)/4$, where $\Omega = \sqrt{\omega_0^{2} 
- D^{2}}$. This explains the appearance of the two distinguished time scales 
$t$ and $\tau$. The exponential Ansatz (\ref{expAnsatz}) was introduced in 
\cite{BON1} motivated by the failure of the usual expansion of $\rho$
in power series of $\varepsilon$ for the particular model considered there.
For that model, an algebraic Ansatz yields a vertical bifurcating branch to
all orders in $\varepsilon$. In other models where the unknown $\rho$
is everywhere non-negative, such an exponential Ansatz yields an 
asymptotic expansion (in $\varepsilon$) with larger 
domain of validity than a purely algebraic Ansatz~\cite{LIN}.

Inserting (\ref{expAnsatz}) and (\ref{tau}) into the governing equations 
(\ref{5})-(\ref{8}), we obtain the hierarchy (3.5a)-(3.7b) of \cite{BNS}:
\begin{eqnarray}
{\cal L}\sigma_1 \equiv (\partial_{t}-D\partial_{\theta}^{2}+\omega
\partial_{\theta}) \sigma_1 - K_{c} \mbox{Re}\, e^{-i\theta} 
\langle e^{i\theta^{\prime}}, \sigma_1 \rangle = 0,\nonumber\\
\label{11}\\
   \int_{0}^{2 \pi} \sigma_{1} d \theta = 0,\nonumber
\end{eqnarray}

\begin{eqnarray}
      {\cal L} \left( \sigma_{2} + \frac{\sigma_{1}^2}{2} \right)
	 = - K_c \partial_{\theta} \left\{ \sigma_{1} \mbox{Im} \ 
e^{-i \theta} \langle e^{i \theta'},\sigma_{1} \rangle  
\right\} ,
\nonumber\\
\label{sigma2}\\
     \int_{0}^{2 \pi} \left( \sigma_{2} + \frac{\sigma_{1}^2}{2}\right) 
d\theta = 0,
\nonumber
\end{eqnarray}

\begin{eqnarray}
      {\cal L} \left( \sigma_{3} + \sigma_{1} \sigma_{2} +
	   \frac{\sigma_{1}^{2}}{6} \right) = - K_c \,\partial_{\theta} \left\{ 
   \left(\sigma_{2} + \frac{\sigma_{1}^{2}}{2} \right)\,\mbox{Im} \ 
e^{-i\theta} \langle e^{i\theta'},\sigma_{1} \rangle \right.\quad\quad
\nonumber\\
\left. + \sigma_{1} \mbox{Im} \ e^{-i\theta} \langle 
e^{i\theta'},\sigma_{2} + \frac{\sigma_{1}^{2}}{2}\rangle \right\}    
- K_{2} [\partial_{\tau} \sigma_{1} + \partial_{\theta} \mbox{Im} \ e^{-i\theta} \langle e^{-i\theta'},
\sigma_{1}\rangle ] , 
\nonumber\\
 \label{sigma3}\\
\int_{0}^{2 \pi} \left( \sigma_{3} + \sigma_{1} \sigma_{2} +
	   \frac{\sigma_{1}^{2}}{6} \right) d\theta = 0.
\nonumber
\end{eqnarray}
Here we have defined the following scalar product~\cite{BNS}
\begin{equation}
     \langle\alpha(\theta,\omega),\beta(\theta,\omega)\rangle =
   \frac{1}{2\pi}\int_{0}^{2\pi}\int_{-\infty}^{+\infty}
  \alpha(\theta,\omega)\beta(\theta,\omega)g(\omega)d\omega d\theta.\label{14}
\end{equation}

The solution of the homogeneous linear equation (\ref{11}) is a linear 
combination of $\mu_l e^{i\Omega t}$, $l=1,2$ and the complex conjugates
of these terms [the $\mu_l$ are given by (\ref{eigenf})]: 
\begin{equation}
     \sigma_{1} = \frac{A_{+}(\tau)}{D + i(\Omega + \omega)}e^{i(\Omega
  t + \theta)} +cc + \frac{A_{-}(\tau)}{D + i(\Omega - \omega)}
		     e^{i(\Omega t - \theta)} + cc,\label{9}
\end{equation}
where $\Omega^2 = \omega_{0}^2 - D^2$ and $cc$ denotes the complex
conjugate of the preceding term (in \cite{BNS} there was $A_{-} \equiv 0$, 
$A_{+} \equiv A$. Thus two terms were missing). This value of $\sigma_{1}$ 
has also zero mean, as a function of $\theta$. Insertion of this equation in 
(\ref{sigma2}) and (\ref{sigma3}) yields 
\begin{eqnarray}
  {\cal L}(\sigma_{2} + \frac{\sigma_{1}^{2}}{2}) = 2 e^{2i\Omega t}
    \left( \frac{A_{+}^{2} e^{2i\theta}}{D + i(\Omega +\omega)}+
    \frac{A_{-}^{2} e^{-2i\theta}}{D + i(\Omega - \omega)}\right) + cc
	\nonumber\\    
+ 4 e^{2i\theta} A_{+}\overline{A}_{-}
     \frac{D + i \omega}{(D+ i\omega)^{2} + \Omega^{2}} + cc ,\label{10}
\end{eqnarray}
from which
\begin{eqnarray}
     \sigma_{2}+\frac{\sigma_{1}^{2}}{2}=
  \frac{2(D+i\omega)A_{+}\overline{A}_{-}}{(2D+i\omega)[(D
      +i\omega)^{2}+\Omega^{2}]}
	   e^{2i\theta} + cc\nonumber\\
	+ \frac{A_{+}^{2}e^{2i(\Omega t+\theta)}}
      {[D+i(\Omega+\omega)][2D+i(\Omega+\omega)]}+ cc\nonumber\\
    + \frac{A_{-}^{2}e^{2i(\Omega t-\theta)}}{[D+i(\Omega-\omega)]
       [2D+i(\Omega-\omega)]} + cc ,\label{12}
\end{eqnarray}
which has also zero mean, as required.
After lengthy but rather elementary calculations to evaluate the
right-hand side of (\ref{sigma3}), this equation 
takes on the form
\begin{equation}
 {\cal L}(\sigma_{3}+\sigma_{1}\sigma_{2}+\frac{\sigma_{1}^{3}}{6})=
      Q_{+}(\tau, \omega) e^{i(\Omega t+\theta)} + cc +
	Q_{-}(\tau, \omega) e^{i(\Omega t-\theta)} + cc,\label{14bis}
\end{equation}
where only the terms that may be resonant have been kept. It is natural to 
look for a solution of the form
\begin{equation}
       \sigma_{3}+\sigma_{1}\sigma_{2}+\frac{\sigma_{1}^{3}}{6}=
  P_{+} e^{i(\Omega t+\theta)} + cc + P_{-} e^{i(\Omega t-\theta)}
+ cc.\label{15}
\end{equation}
We determine $P_{\pm}$ by substitution of (\ref{15}) into (\ref{14bis}),
\begin{eqnarray}
[D + i(\Omega \pm \omega)] P_{\pm} - {K_{c}\over 2}\,\langle 1,P_{\pm}\rangle
= Q_{\pm} .\nonumber 
\end{eqnarray}
Then we can solve for $P_{\pm}$:
\begin{eqnarray}
 P_{\pm} = {K_{c}\langle 1,P_{\pm}\rangle\over 
2[D + i(\Omega \pm \omega)]} + {Q_{\pm}\over 
D + i(\Omega \pm \omega)}\, .\label{P} 
\end{eqnarray}
>From (\ref{autov}) and the reflection symmetry of $g(\omega)$, we know that 
${1\over 2} K_c \langle 1,1/[D + i(\Omega \pm \omega)]\rangle = 1$, so that
the scalar product of 1 with (\ref{P}) produces the following
{\it non-resonance} conditions:
\begin{eqnarray}
     \langle \frac{Q_{+}}{D+i(\Omega+\omega)} \rangle=0   \nonumber  \\
     \langle \frac{Q_{-}}{D+i(\Omega-\omega)} \rangle=0,\label{16}
\end{eqnarray}
where we set
\begin{equation}
    \langle \alpha(\omega) \rangle =
  \int_{-\infty}^{+\infty}\alpha(\omega)g(\omega)d\omega.\label{17}
\end{equation}
The zero mean condition is also satisfied automatically. Some more
tedious calculations lead finally to two nonlinear coupled ordinary
differential equations for $A_{+}(\tau), A_{-}(\tau)$:
\begin{eqnarray}
   \dot{A}_{+}=\alpha A_{+} - (\beta|A_{-}|^{2}+\gamma |A_{+}|^{2})\,
A_{+}, \nonumber\\
\label{19}\\
   \dot{A}_{-}=\alpha A_{-} - (\beta|A_{+}|^{2}+\gamma|A_{-}|^{2})\,
A_{-}, \nonumber
\end{eqnarray}
where \, $\dot{} = d/d\tau$, and 
\begin{eqnarray}
\alpha = \frac{1}{4}-\frac{iD}{4 \Omega}\, ,\nonumber\\
\beta = \frac{D+i{D^{2}+\omega_{0}^{2}\over\Omega}}{K_{2}\, (4D^{2}+
\omega_{0}^{2})}\, , \nonumber\\
\gamma = \frac{2(3D^{2}+4\omega_{0}^{2})+iD\, {3D^{2}+2\omega_{0}^{2}
\over\Omega}}{D K_{2}\, (9D^{2}+16\omega_{0}^{2})} \, . \label{constants}
\end{eqnarray}
This result favorably agrees with that of ~\cite{BNS} when we set
$A_{+} \equiv A,\, A_{-} \equiv 0$. The needed stability analysis is,
consequently, a little more involved than that in ~\cite{BNS}. Let us 
define the new variables 
\begin{equation}
    u=|A_{+}|^{2} + |A_{-}|^{2}, \ \ \   v=|A_{+}|^{2} - |A_{-}|^{2}.
\label{20}
\end{equation}
By using (\ref{19}), we obtain the following system for $u$ and $v$:
$$
   \dot{u}=2 \ \mbox{Re} \ \alpha \ u - \mbox{Re} (\gamma+\beta) \ u^{2} -
       \mbox{Re} (\gamma-\beta) \ v^{2} ,
$$
\begin{equation}
   \dot{v}=2 \ \mbox{Re} \ \alpha \ v - 2 \ \mbox{Re} \ \gamma \ u v.\label{21}
\end{equation}
Clearly, $u = v$ or $u = -v$ correspond to {\it traveling wave} (TW) solutions,
while $v \equiv 0$ corresponds to {\it standing wave} (SW) solutions. The phase
portrait corresponding to $\alpha$, $\beta$ and $\gamma$ of (\ref{19}) is 
easily found (see Fig.~2), and the explicit solutions are (up to, possibly, a 
constant phase shift)
\begin{equation}
       A_{+} (\tau) = \sqrt{\frac{\mbox{Re} \ \alpha}{\mbox{Re} \ \gamma}}
	   \ e^{i \mu \tau}, \ \   A_{-} (\tau) \equiv 0,  \ \ \
		\mu = \mbox{Im} \ \alpha - \frac{\mbox{Im} \
	    \gamma}{\mbox{Re} \  \gamma} \, \mbox{Re} \ \alpha \label{22}
\end{equation}
(or $A_{+}(\tau) \equiv 0$ and $A_{-}(\tau)$ as $A_{+}(\tau)$ above) in 
case of TW solutions, and
\begin{equation}
      A_{+} (\tau) = A_{-} (\tau) = \sqrt{\frac{\mbox{Re} \
	    \alpha}{\mbox{Re} \ (\gamma + \beta)}} \ e^{i \nu \tau},
     \ \ \   \nu = \mbox{Im} \ \alpha - \frac{\mbox{Im} \ (\gamma +
	 \beta)}{\mbox{Re} \ (\gamma + \beta)} \,\mbox{Re} \ \alpha \label{23}
\end{equation}
in case of SW solutions. Notice that both SW and TW bifurcate 
supercritically with $\|r_{SW}\|/r_{TW}>1$, as indicated in 
Fig.~3: Re$(\beta +\gamma)$ and Re$\gamma$ are both positive when 
$K_2 = 1$; whereas the square roots in (\ref{22}) and (\ref{23}) become 
pure imaginary if $K_2=-1$. This indicates that the bifurcating branches 
cannot be subcritical. From the phase portrait corresponding to (\ref{19}), 
it follows that the SWs are always {\it globally stable}, while the TWs are 
{\it unstable}. Such result was pointed out in \cite{CRAW}, following
completely different methods, while in \cite{BNS} the analysis was restricted
to the case $u^2 = v^2$, and thus the TWs were erroneously found to
be stable. 



\section{Multiscale analysis near the tricritical point}

   Asymptotic analysis near the tricritical point, $P = (K/D = 4, \omega_{0}/D
= 1)$ in Fig.1, leads to the introduction of a {\it third} time-scale. In
fact, near such a point,
\begin{equation}
    K = K_{c} + K_{2} \varepsilon^2 + O(\varepsilon^3),  \ \ \
     \omega_{0} = \omega_{0c} + \omega_{2} \varepsilon^2 + O(\varepsilon^3)
     \ \ \     (K_{c} = 4 D, \  \omega_{0c} = D),\label{24}
\end{equation}
and
\begin{equation}
       \lambda_{\pm} = -D + \frac{K}{4} \pm \frac{1}{4} \sqrt{K^2 - 16
		   \omega_{0}^2}
	   \approx \frac{K_{2}}{4} \varepsilon^2 \pm \frac{\varepsilon}{4}
		 \sqrt{8 D (K_{2} - 4 \omega_{2})}.\label{25}
\end{equation}
This shows that, besides the basic time-scale (which is denoted by $t$), and
the {\it slow} time $\tau = \varepsilon^2 t$ (as in \cite{BNS}), an {\it
intermediate} scale, say $T = \varepsilon t$, appears. Compare
\begin{equation}
	 e^{\lambda_{\pm} t} \sim \exp \left\{ \frac{K_{2}}{4} \tau \pm
	\frac{\sqrt{8 D (K_{2} - 4 \omega_{2})}}{4} T \right\}\, \label{26}
\end{equation}
with (\ref{expfactor}) above. Consequently, the slightly different Ansatz
\begin{equation}
      \rho (\theta, t; \omega; \varepsilon)
	 = \frac{1}{2 \pi} \exp \left\{ \sum_{j = 1}^{4} \varepsilon^j
	  \sigma_{j} (\theta, t, T, \tau) + O(\varepsilon^5) \right\}\, \label{27}
\end{equation}
is needed. Inserting (\ref{24}) and (\ref{27}) 
into the governing equations (\ref{5})-(\ref{8}) leads to the hierarchy 
below, instead of (\ref{11})-(\ref{sigma3}):
\begin{eqnarray}
  {\cal L} \sigma_{1} = (\partial_{t}-D \partial_{\theta}^2 + \omega 
\partial_{\theta})	\sigma_{1}
    + 4 D \partial_{\theta} \left\{ \mbox{Im} \ e^{-i \theta} \langle e^{i
	 \theta'}, \sigma_{1} \rangle \right\} = 0,\nonumber\\
\label{28}\\
   \int_{0}^{2 \pi} \sigma_{1} d \theta = 0,\nonumber
\end{eqnarray}

\begin{eqnarray}
      {\cal L} \left( \sigma_{2} + \frac{\sigma_{1}^2}{2} \right)
	 = -4 D \partial_{\theta} \left\{ \sigma_{1} \mbox{Im} \ 
e^{-i \theta} \langle e^{i \theta'},\sigma_{1} \rangle  
\right\}  - \partial_{T} \sigma_{1},
\nonumber\\
\label{29}\\
     \int_{0}^{2 \pi} \left( \sigma_{2} + \frac{\sigma_{1}^2}{2}\right) 
d\theta = 0,
\nonumber
\end{eqnarray}

\begin{eqnarray}
      {\cal L} \left( \sigma_{3} + \sigma_{1} \sigma_{2} +
	   \frac{\sigma_{1}^{2}}{6} \right) = -4 D \,\partial_{\theta} \left\{ 
 \sigma_{1} \mbox{Im} \ e^{-i\theta} \langle 
e^{i\theta'},\sigma_{2} + \frac{\sigma_{1}^{2}}{2}\rangle  + \left(\sigma_{2} 
+ \frac{\sigma_{1}^{2}}{2} \right)\mbox{Im} \ e^{-i\theta} \langle 
e^{i\theta'},\sigma_{1} \rangle \right. \nonumber\\	    
\left. + \omega_{2}\, \mbox{Im} \ e^{-i \theta} \langle e^{i \theta'},
  \sigma_{1} \rangle'  \right\}
- K_{2} \partial_{\theta} \mbox{Im} \ e^{-i\theta} \langle e^{-i\theta'},
\sigma_{1}\rangle - \partial_{\tau} \sigma_{1} 
- \partial_{T} \left( \sigma_{2} + \frac{\sigma_{1}^{2}}{2} \right) , 
\nonumber\\
 \label{29bis}\\
\int_{0}^{2 \pi} \left( \sigma_{3} + \sigma_{1} \sigma_{2} +
	   \frac{\sigma_{1}^{2}}{6} \right) d\theta = 0,
\nonumber
\end{eqnarray}

\begin{eqnarray}
   {\cal L} \left( \sigma_{4} + \sigma_{1} \sigma_{3} + \frac{\sigma_{2}^2}{2}
    + \frac{\sigma_{1}^2 \sigma_{2}}{2} + \frac{\sigma_{1}^4}{4!} \right)
     = -4 D \partial_{\theta} \left\{ \sigma_{1} \mbox{Im} \ e^{-i \theta}
      \langle e^{-i \theta'}, \sigma_{3} + \sigma_{1} \sigma_{2}
	+ \frac{\sigma_{1}^3}{6} \rangle \right.
\nonumber\\   
\left.	 + \left( \sigma_{3} + \sigma_{1} \sigma_{2} +
  \frac{\sigma_{1}^3}{6} \right) \mbox{Im} \ e^{-i \theta} \langle e^{i
  \theta'}, \sigma_{1} \rangle + \omega_{2} \mbox{Im} \ e^{-i \theta}
  \langle e^{i \theta'}, \sigma_{2} + \frac{\sigma_{1}^2}{2} \rangle'
	  \right.\nonumber\\
      \left.   + \omega_{2} \sigma_{1} \mbox{Im} \ e^{i \theta} \langle
	e^{i \theta'}, \sigma_{1} \rangle' + \left( \sigma_{2} +
  \frac{\sigma_{1}^2}{2} \right) \mbox{Im} \ e^{-i \theta} \langle e^{i
    \theta'}, \sigma_{2} + \frac{\sigma_{1}^2}{2} \rangle  \right\}
\nonumber\\
- K_{2}\, \partial_{\theta} \left\{ \mbox{Im} \ e^{-i \theta} \langle e^{i \theta'}, 
\sigma_{2} + \frac{\sigma_{1}^2}{2} \rangle + \sigma_{1} \mbox{Im} \ e^{-i
\theta} \langle e^{i \theta'}, \sigma_{1} \rangle \right\}
\nonumber\\
- \partial_{\tau} \left( \sigma_{2} + \frac{\sigma_{1}^{2}}{2} \right)	 
- \partial_{T} \left( \sigma_{3} + \sigma_{1} \sigma_{2} +
		    \frac{\sigma_{1}^2}{2} \right),
\nonumber\\
\label{30}\\
       \int_{0}^{2 \pi} \left( \sigma_{4} + \sigma_{1} \sigma_{3}
	   + \frac{\sigma_{2}^2}{2} + \frac{\sigma_{1}^2 \sigma_{2}}{2}
	    + \frac{\sigma_{1}^4}{4!} \right) d \theta = 0.
\nonumber
\end{eqnarray}
Here
\begin{equation}
    \langle \alpha(\theta, \omega), \beta(\theta, \omega) \rangle'
	= \frac{1}{2 \pi} \int_{0}^{2 \pi} \int_{- \infty}^{+ \infty}
      \alpha(\theta, \omega) \beta(\theta, \omega) g_{\omega_{0}}'
	 (\omega) d \theta d \omega,\label{32}
\end{equation}
where
\begin{equation}
       g_{\omega_{0}}' (\omega) = \frac{1}{2} \left[ \delta'(\omega +
	 \omega_{0}) - \delta'(\omega - \omega_{0}) \right]
	   \ \ \ \ \	( \omega_{0} = \omega_{0c} = D).\label{33}
\end{equation}
The solution of the homogeneous equation (\ref{28}) for $\sigma_{1}$ is 
immediately found [$\Omega = 0$ in (\ref{9})]:
\begin{equation}
	\sigma_{1} = \frac{A(T,\tau)}{D + i \omega} e^{i \theta} + cc, \label{34}
\end{equation}
plus terms which decay exponentially on the fast time scale, $t$, and which
we will systematically omit. Inserting this into equation (\ref{29}), we obtain
\begin{equation}
       {\cal L} \left( \sigma_{2} + \frac{\sigma_{1}^2}{2} \right)
	  = - \frac{A_{T}}{D + i \omega} e^{i \theta} + cc
	  + \frac{2 A^2}{D + i \omega} e^{2 i \theta} + cc,\label{35}
\end{equation}
wherefrom
\begin{equation}
     \sigma_{2} + \frac{\sigma_{1}^2}{2} = - \frac{A_{T}}{(D + i \omega)^2}
	e^{i \theta} + cc + \frac{A^2}{(D + i \omega)(2 D + i \omega)}
	 e^{2 i \theta} + cc + \frac{B(T, \tau)}{D + i \omega} e^{i
	     \theta} + cc,\label{36}
\end{equation}
and hence $\sigma_{2}$. Note that the term containing $B(T,\tau)$ is the
solution of the homogeneous equation associated to $\cal L$ [cf. (\ref{34})].
Proceeding in a similar way, we obtain
\begin{eqnarray}
      \sigma_{3} + \sigma_{1} \sigma_{2} + \frac{\sigma_{1}^3}{6}
	 = \left[ \frac{K_{2} - 4 \omega_{2}}{4 D (D + i \omega)} A
     - \frac{B_{T}}{(D + i \omega)^2} + \frac{A_{TT}}{(D + i \omega)^3}
	  - \frac{A_{\tau}}{(D + i \omega)^2}	+ \frac{C(T, \tau)}{D + i \omega} \right.
 \nonumber\\
\left.    
- \frac{A |A|^{2}}{(D + i \omega)^{2} (2 D + i \omega)} \right]\, e^{i \theta} 
+ cc + \frac{2 A B - A A_{T} \left(\frac{1}{D + i \omega} + \frac{1}{2 D
  + i \omega} \right)}{(D + i \omega)(2 D + i \omega)} e^{2 i \theta}
\nonumber\\
 + cc  + \frac{A^3 e^{3 i \theta}}{(D + i \omega)(2 D + i \omega)(3 D + i
   \omega)} + cc , \label{37}
\end{eqnarray}
where $C\equiv C(T,\tau)$ has a meaning similar to that of $A$ and $B$. 
>From this we obtain $\sigma_{3}$, and finally, from (\ref{30}), $\sigma_{4}$. 
To obtain the leading order approximation, we only need to determine $A(T,
\tau)$. Now, (\ref{37}) holds provided that the {\it nonresonance condition} 
(needed to remove secular terms)
\begin{equation}
       \langle \frac{1}{D + i \omega}, P(\omega, T, \tau) \rangle = 0\label{38}
\end{equation}
holds, where $P(\omega, T, \tau)$ denotes the coefficient of $e^{i \theta}$
on the right-hand side of (\ref{29bis}). Equation (\ref{38}) turns out to be 
the ``complex Duffing equation''
\begin{equation}
    A_{TT} - \frac{D}{2} (K_{2} - 4 \omega_{2}) A - \frac{2}{5} |A|^2 A = 0.
\label{39}
\end{equation}
Such equation, however, is {\it not} sufficient to determine $A$, in view of
the {\it two} time scales on which $A$ depends. The nonresonance condition for
$\sigma_{4}$, i.e. an equation like that in (\ref{38}) where $P(\omega,T,\tau)$
now denotes the coefficient of $e^{i \vartheta}$ on the right-hand side of
(\ref{30}), is the ``linearized inhomogeneous Duffing equation"
\begin{eqnarray}
      B_{TT} - \frac{D}{2} (K_{2} - 4 \omega_{2}) B - \frac{2}{5} (A^2
	  \overline{B} + 2 |A|^2 B) 
  = - 2 A_{T \tau} + \frac{K_{2}}{2} A_{T} \nonumber\\
- \frac{\left(|A|^2 A \right)_{T}}{5 D} - \frac{23}{25 D} |A|^2 A_{T},
\label{40}
\end{eqnarray}
where an overbar denotes taking the complex conjugate.

Equations (\ref{39}) and (\ref{40}) could be analyzed directly, e.g. by 
extending the Kuzmak-Luke method (see \cite{Cole}, Section 4.4), to find the 
bifurcating solutions in the vicinity of the tricritical point and their 
stability. However, we can take advantage from the already existing, rather 
comprehensive theory of amplitude equations for systems invariant under
the $O(2)$ group of rotations ($\theta\to \theta+\varphi$) and reflections 
($\theta\to -\theta$, $\omega\to -\omega$) developed by Dangelmayr and Knobloch 
in \cite{DK}. Our nonlinear Fokker-Planck problem has this symmetry, therefore
the normal form near the tricritical point (a Takens-Bogdanov bifurcation)
should be the same one that Dangelmayr and Knobloch studied. 
Equations (\ref{39}) and (\ref{40}) in fact can be used to reconstruct the 
scaled ``normal form'':
\begin{equation}
   U'' - \varepsilon [c_{1} U' + c_{2} (\overline{U} U' + U \overline{U}')\, U
	  + c_{3} |U|^2 U'] - (c_{4} + c_{5} |U|^2) U = O(\varepsilon^2),
		   \ \ \ \   ' = \frac{d}{d T}\, ,\label{41}
\end{equation}
studied by Dangelmayr and Knobloch in \cite{DK} [cf.\ their equations (3.3), 
p.\ 2480]; recall that $T=\varepsilon t$ is the slow scale. Setting
\begin{equation}
      U= A(T, \tau) + \varepsilon B(T, \tau) \equiv A(T, \varepsilon T)
	      + \varepsilon B(T, \varepsilon T)\label{42}
\end{equation}
in (\ref{41}), we obtain equations for $A$ and $B$ which are of the same form
as (\ref{39}) and (\ref{40}). We can then identify the parameters termed
$\mu$, $\nu$, $A$, $C$, $D$ in \cite{DK}, and thus $M= 2 C +D$ there, with our
quantities
\begin{equation}
    \frac{D}{2} (K_{2} - 4 \omega_{2}), \ \ \	\frac{K_{2}}{2}, \ \ \
     \frac{2}{5}, \ \ \  - \frac{1}{5 D}, \ \ \  - \frac{28}{25 D}, \ \ \
       \mbox{and} \ \  - \frac{38}{25 D},\label{43}
\end{equation}
respectively. With these identifications, Equation (\ref{41}) becomes
\begin{eqnarray}
    U_{TT} - \frac{D}{2} (K_{2} - 4 \omega_{2}) U - \frac{2}{5} |U|^2 U
	 = \varepsilon\,\left(\frac{K_{2}}{2}  U_{T} - \frac{23}{25 D}
	    |U|^2 U_{T} - \frac{1}{5 D} (|U|^2 U)_{T}\right) + O(\varepsilon^2).
\label{44}
\end{eqnarray}
Note that $-2 A_{T \tau} \varepsilon = O(\varepsilon^2)$. The general analysis
developed in \cite{DK} for equation (\ref{41}) can be used
for the present case, equation (\ref{44}) [cf.~\cite{DK}, equation (3.3)]. We 
make the substitution
\begin{equation}
      U(T;\varepsilon) = R(T; \varepsilon) e^{i \phi(T; \varepsilon)}
\label{45}
\end{equation}
in equation (\ref{44}), separate real and imaginary parts, and then obtain the
perturbed Hamiltonian system
\begin{eqnarray}
    R_{TT} + \frac{\partial V}{\partial R} = \varepsilon \left( \frac{K_{2}}{2} -
	      \frac{38}{25 D} R^2 \right) R_{T} , \nonumber\\
\label{46}\\
  L_{T} = \varepsilon \left( \frac{K_{2}}{2} - \frac{28}{25 D} R^2 \right) L,
\nonumber
\end{eqnarray}
where
\begin{equation}
	L= R^2 \phi_{T}\label{47}
\end{equation}
is the angular momentum, and
\begin{equation}
       V \equiv V(R) = \frac{L^2}{2 R^2} - \frac{D}{4} \left( K_{2} - 4
	      \omega_{2} \right) R^2 - \frac{R^4}{10} \label{48}
\end{equation}
is the potential. This system may have the following special solutions (whose
stability properties are also pointed out here):
\begin{itemize}
\item[$(i)$]
  The {\it trivial solution}, $L = 0$, $R = 0$, which corresponds to the
incoherent probability density, $\rho = 1/2 \pi$. Such solution is stable 
for $K_2 <0$ if $\omega_2 > 0$ and for $(K_{2} - 4 \omega_{2}) < 0$
if $\omega_2 < 0$.
\item[$(ii)$]
  The {\it steady-states} (SS), $L = 0$, $R = R_{0} = \sqrt{5 D \left( 
\omega_{2} - \frac{K_{2}}{4} \right)} > 0$, which exists provided that 
$\omega_{2} > K_{2}/4$. This solution is always unstable.
\item[$(iii)$]
  The {\it traveling waves} (TW), $L = L_{0} = R_{0}^2 \sqrt{2 D \left(
\omega_{2} - \frac{19}{56} K_{2} \right)} > 0$, $R = R_{0} = \frac{5}{2}
\sqrt{\frac{D K_{2}}{14}} > 0$, which exist provided that $K_{2} > 0$ and
$\omega_{2} > 19 K_{2}/56$; these solutions bifurcate from the trivial solution
at $K_{2} = \omega_{2} = 0$. When $\omega_{2} = 19 K_{2}/56$, the branch of
TWs merges with the steady-state solution branch. This solution is
always unstable.
\item[$(iv)$]
  The {\it standing waves} (SW), $L = 0$, $R = R(T)$ periodic. Such solutions
have been found explicitly in Section 5.1 of \cite{DK}. The SWs branch off 
the trivial solution at $K_{2} = \omega_{2} = 0$, exist for $\omega_{2} > 11 
K_{2}/19 > 0$, and terminate by merging with a homoclinic orbit of the 
steady-state $(ii)$ on the line $\omega_{2} = 11 K_{2}/19$ [see 
equation (5.8) of \cite{DK}]. This solution is always stable.
\end{itemize}

   All these results are depicted in Fig.~3 below, which corresponds to Fig.~4,
IV-, in the general classification (stability diagrams) reported in \cite{DK},
p.266.


\noindent  In Fig. 4 below, the bifurcation diagram relevant to the 
present problem with $\omega_{2} > 0$ is given (cf. Fig.5, IV-, in \cite{DK}, 
p.267).


Note that the modulated wave solutions (in the terminology of \cite{DK}), i.e.
with both $L$ and $R$ periodic functions, in general with different periods, do
{\it not} appear in the problem studied in the present paper.

   In closing, observe that, to the leading order, equation (\ref{27}) yields
\begin{equation}
   \rho(\theta, t; \omega; \varepsilon) \sim \frac{1}{2 \pi} \left[ 1 +
     \varepsilon \frac{R e^{i (\phi + \theta)}}{D + i \omega} + cc \right],
\label{49}
\end{equation}
and hence, from (\ref{7}),
\begin{equation}
      r e^{i \psi} \sim \varepsilon \frac{R}{2 D}e^{-i \phi}.\label{50}
\end{equation}
It follows that
\begin{equation}
      r \sim \varepsilon \frac{R}{2 D},  \qquad  \psi \sim - \phi,\label{51}
\end{equation}
which shows that, essentially, the solution $U(T;\varepsilon)$ to equation
(\ref{44}) coincides with the conjugate of the complex order parameter 
[defined by (\ref{7})]. For this reason, in Fig.~4 the ordinate can be 
either $R$ or $r$. In Fig.~5, we depicted the global bifurcation diagram 
which completes the analogous one given in \cite{BNS}, cf. Fig.~5 there.


\section{Numerical results}

The goal of this section is to give numerical evidence of the theoretical
results obtained thus far. To perform this task, we have integrated the 
stochastic Eq.~(\ref{4}) by a first-order Euler method with a time step $\Delta 
t = 0.005$. In all our simulations a population of $N=50,000$ has been chosen, 
which is large enough to neglect finite-size effects. 

The interesting region in the space of parameters is located
above the tricritical point $(K/D =4, \omega_0/D=1)$. To explore this
region and without loss of generality, we have kept fixed the strength of the
noise to $D=1$. 
Then we have set $\omega_0=2$, and we have swept the phase diagram by moving 
the coupling constant, $K$, thereby finding different behavior according to 
the results of the previous sections. Consistently with the figures 
depicted above, we have considered only values $K>4$, for which the 
incoherent solution $\rho \equiv 1/ 2\pi$ is unstable. For these values of
$K$, the (partially) synchronized SW states bifurcate supercritically and
are stable until the SW branch disappears. In this section 
we define the order parameter (\ref{3}) or (\ref{7}) in such a way that 
$r(t)\in [-1,1]$ and that the phase does not experiences jumps as it 
increases past odd integer multiples of $\pi$. Then the order parameter 
which we should use to compare with the results of previous
Sections is $|r(t)|\geq 0$.  

Let start the discussion considering $K=5.2$ . In Fig.\ 6, we can see that, 
after a short transient, the order parameter $|r(t)|$ reaches a stable state 
characterized by time-periodic oscillations of large amplitude. Clearly, 
this value of the coupling constant belongs to the domain of the SW solution. 
This periodic behavior is found as soon as $K$ becomes larger than 4, but 
near the critical point the frequency of the oscillations
is very high (recall that $\omega_0=2$) and their amplitude quite small. 
This is why we do not depict such a behavior in any of the figures. 
Moreover, when $K=5.2$, the Fourier transform of the order parameter exhibits 
a large peak at a {\em nonzero} frequency, which corresponds to a relaxation 
oscillation. This peak slowly fades out as $K$ decreases down towards $K=4$ 
(near the bifurcation point the oscillation becomes sinusoidal). 


The opposite behavior is found for larger values of $K$. 
In comparison with the last figure now the amplitude of the oscillations 
increases while the frequency decreases in a nontrivial way with the coupling
constant as we can see in Fig. 7 for $K=6$. The system still remains in the
domain where the standing waves are stable.


According to the theory, the SW solution
should merge with the SS solution for values of $K$ large enough. Indeed,
this is what we observe in Fig. 8. In this case for $K=7$ the
order parameter grows exponentially fast from the initial incoherent
solution to the time-independent partially synchronized stationary state. 
The conjectured global bifurcation diagram of Fig.~5 suggests that there
may be a region where the SW and the partially synchronized stationary
solution are both stable. In order to detect the presence of bistability,
it is more convenient to use a deterministic numerical method to solve 
the nonlinear Fokker-Planck equation. In fact, the Monte Carlo simulation 
averages over realizations of the noise. Then different realizations may  
go to different stable solutions in the bistability region, {\em unless we are
rather careful choosing convenient initial conditions within the basin
of attraction of one solution, and a small enough time step}. Then we need
an enormous amount of computing time for a Monte Carlo simulation to 
distinguish the attractor with smaller basin of attraction in the bistability 
region. Thus we have used deterministic numerical simulations (finite 
differences) of the nonlinear Fokker-Planck equation to obtain the results 
reported below, although we have checked that costly Monte Carlo simulations 
also yield the same results in several points of the bifurcation diagram. 
A direct numerical 
simulation of the nonlinear Fokker-Planck equation by finite differences
shows that for sufficiently large $\omega_0$ the region of bistability 
disappears. At $\omega_0=1.5$ we have found a narrow region of bistability 
between SW and SS solutions which is illustrated in Fig.~9. Fig.~9(a) shows
that different initial data evolve either to the SW or to the upper SS
solution for $K=4.95$. Fig.~9(b) illustrates the abrupt transition from a
SW solution to the upper SS solution when $K$ changes from 4.9597 to 4.9598.
When $\omega_0$ is larger, $\omega_0 = 2$ as in Fig.~10, direct simulations
show a smooth transition from SW to SS. This may correspond to having the 
turning point $K_1$ of Fig.~5 close to the end point of the SW branch. 


\section{Summary}

We have used the method of multiple scales to study synchronization to 
oscillatory phases in the mean-field Kuramoto model with a bimodal 
frequency distribution. Near the Hopf bifurcation points our method 
recovers Crawford's results: solution branches of stable standing waves 
(SW) and unstable traveling waves (TW) issue supercritically from the 
incoherent (non-synchronized) state. Near the tricritical point (where 
a line of Hopf bifurcations and a line of partially-synchronized 
stationary states coalesce) our multiple scale method recovers the 
normal form for symmetric Takens-Bogdanov bifurcations studied by 
Dangelmayr and Knobloch. This study allows us to establish that the 
bifurcating branches given by the local analysis of Section II end as 
infinite-period bifurcation solutions. The unstable TW branch terminates
on the SS branch, whereas the SW branch collides with the homoclinic loop
of the SS branch in a global bifurcation of finite amplitude. All 
results obtained in Sections II and III above agree quantitatively, as it
can be shown by asymptotic matching (see Appendix). Furthermore,
there may be an interval of parameter values where SW and 
partially-synchronized stationary solutions are both stable. Brownian
and direct finite-difference simulations (Section IV) confirm these results.

\section{ACKNOWLEDGMENTS}
\label{acknowledgements}

We are indebted to J.~A.\ Acebr\'on for drawing figures 1 to 5, 9 and 10 
and performing direct numerical simulations of the nonlinear Fokker-Planck 
equation.  
We acknowledge financial support from the the Spanish DGICYT
through grant PB95-0296, from the Italian Gruppo Nazionale di Fisica 
Matematica GNFM-CNR, and the EC Human Capital 
and Mobility Programme under contract ERBCHRXCT930413.

\setcounter{equation}{0}
\renewcommand{\theequation}{A.\arabic{equation}}
\section*{Appendix}

   The bifurcation diagrams in Sections II and III agree in the sense that the
corresponding solutions match asymptotically on some overlap domain. For
instance, in case of TW solutions, $A_{+} \neq 0$, $A_{-} \equiv 0$, one
obtains from (\ref{19})
\begin{equation}
   A_{+} e^{i\Omega t} \sim R_{0} \exp\left\{i\Omega t -i \frac{\tau}{\Omega} 
             \left(\frac{D}{4} + \frac{R_{0}^2}{5 K_{2}} \right) \right\}
	   \sim R_{0} \exp \left\{ i T\left[\sqrt{2 D\omega_{2}} -
\frac{K_{2}}{ \sqrt{2 D \omega_{2}}} \left( \frac{D}{4} + \frac{R_{0}^2}{5
K_{2}} \right) \right] \right\},
\label{A1}
\end{equation}
where $R_{0}$ is a constant to be found by asymptotic matching, and
$\tau D/\Omega \sim K_{2} T\sqrt{{D\over 2 \omega_{2}}} = O(1)$, $K_{2} > 0$ 
fixed, as $\omega_{2} \to 0$ from above. On the other hand, near the 
tricritical point, it is shown in Section III that
\begin{equation}
    A \sim \sqrt{\frac{25 D K_{2}}{56}} \exp \left\{ i T \sqrt{2 D
	\left( \omega_{2} - \frac{19}{56} K_{2} \right)} \right\}.\label{A2}
\end{equation}
Let us fix $\omega_{2} > 0$ in this equation and let $K_{2} \to 0$ from
above. Then
\begin{equation}
   \sqrt{2 D \left( \omega_{2} - \frac{19}{56} K_{2} \right)}
       \sim \sqrt{2 D \omega_{2}} - \frac{19}{56}\, K_{2}\, 
        \sqrt{\frac{D}{2\omega_{2}}},\label{A3}
\end{equation}
and inserting the latter into equation (\ref{A2}), asymptotic matching with
equation (\ref{A1}) yields
\begin{equation}
	R_{0} = \sqrt{\frac{25}{56} D K_{2}}.\label{A4}
\end{equation}
The more involved case of the SW branch can be handled in a similar way, 
resorting to the results of reference \cite{DK}.

\begin{figure}
\centerline{\hbox{\psfig{figure=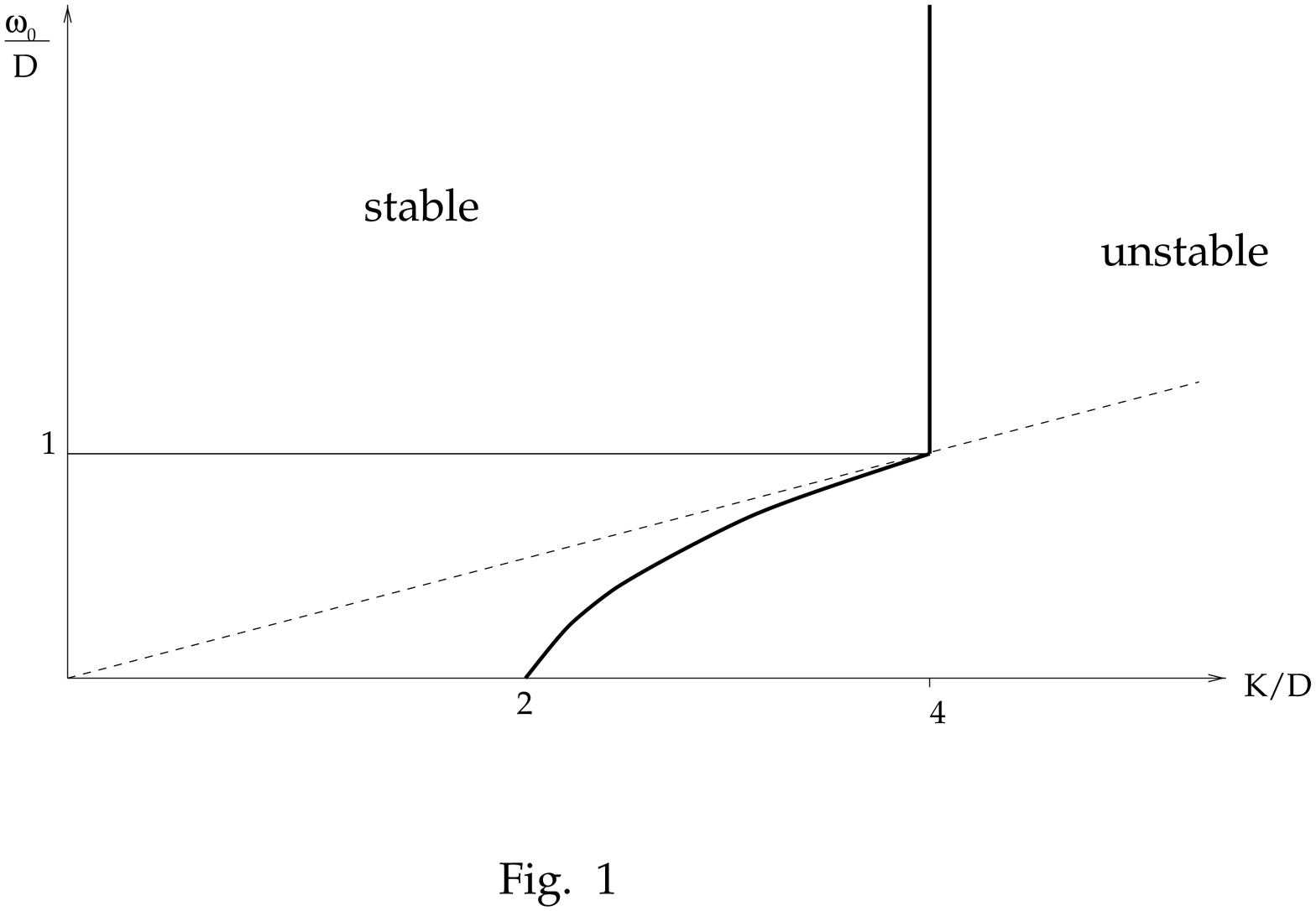,width=9.0cm,angle=-90}}}
\caption{Linear stability diagram for the incoherent solution $\rho_0
= 1/(2\pi)$ and the discrete bimodal frequency distribution, $g(\omega) = 
[\delta(\omega-\omega_0) + \delta(\omega+\omega_0)]/2$ in the parameter 
space $(K/D,\omega_0/D)$. $\rho_0$ is linearly stable to the left of the
lines $K=4D$, $\omega_0 > D$ (where Hopf bifurcations take place) and
$K/(2D) = 1 + \omega_0^2/D^2$, $\omega_0 < D$ (where one eigenvalue of the 
linearized problem becomes zero). To the right of these lines, the 
incoherent solution is linearly unstable. At the tricritical point $K=4D$, 
$\omega_0 = D$, two eigenvalues become simultaneously zero. The dashed line 
separates the region where eigenvalues are real (below the line) from that 
where they are complex conjugate (above the line).
}
\end{figure}

\begin{figure}
\centerline{\hbox{\psfig{figure=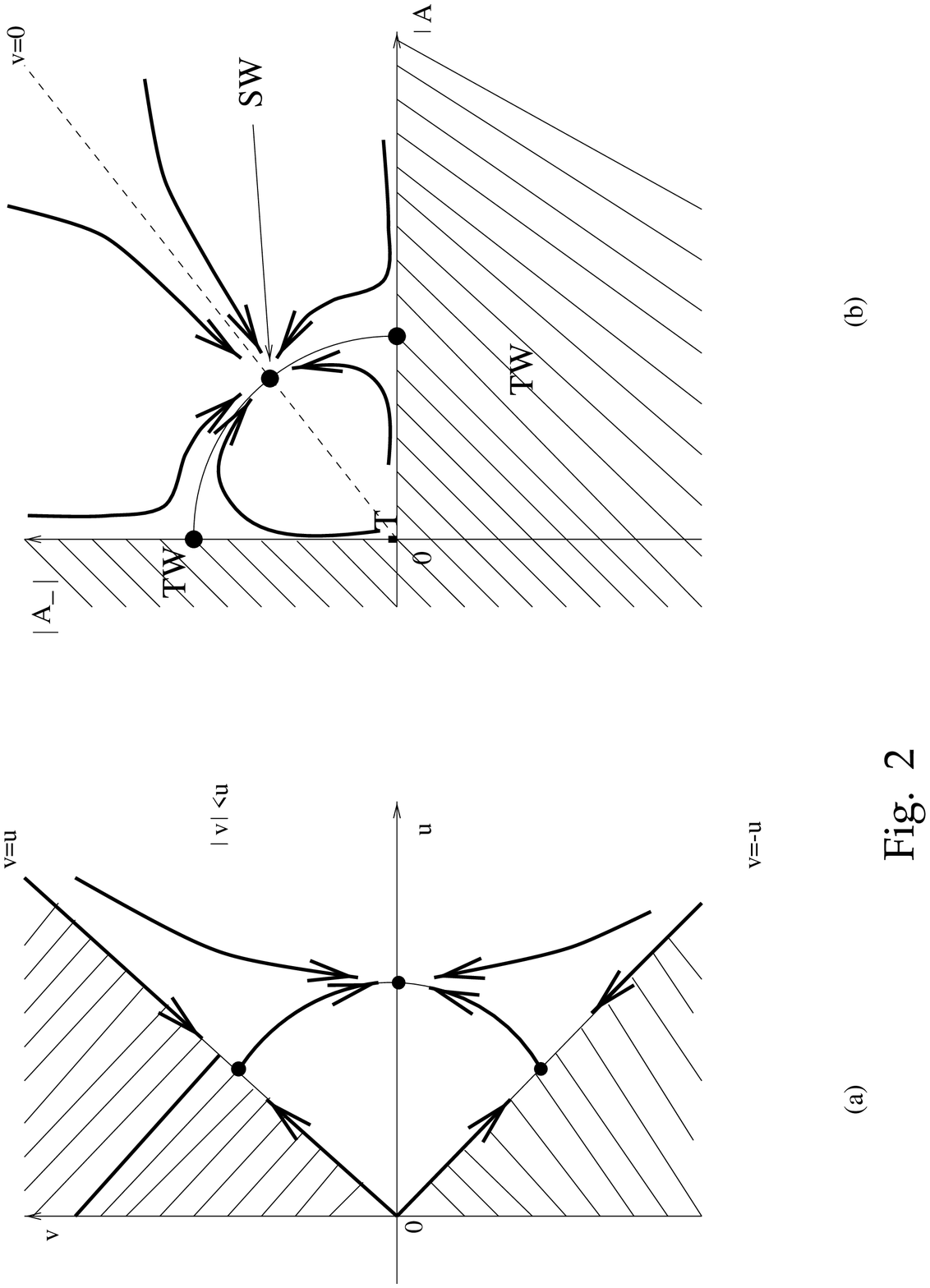,width=9.0cm,angle=-90}}}
\caption{Phase planes (a) $(u,v)$, and (b) $(|A_+|,|A_-|)$ showing the
critical points corresponding to traveling (TW) and standing wave (SW) 
solutions.}
\end{figure}

\begin{figure}
\centerline{\hbox{\psfig{figure=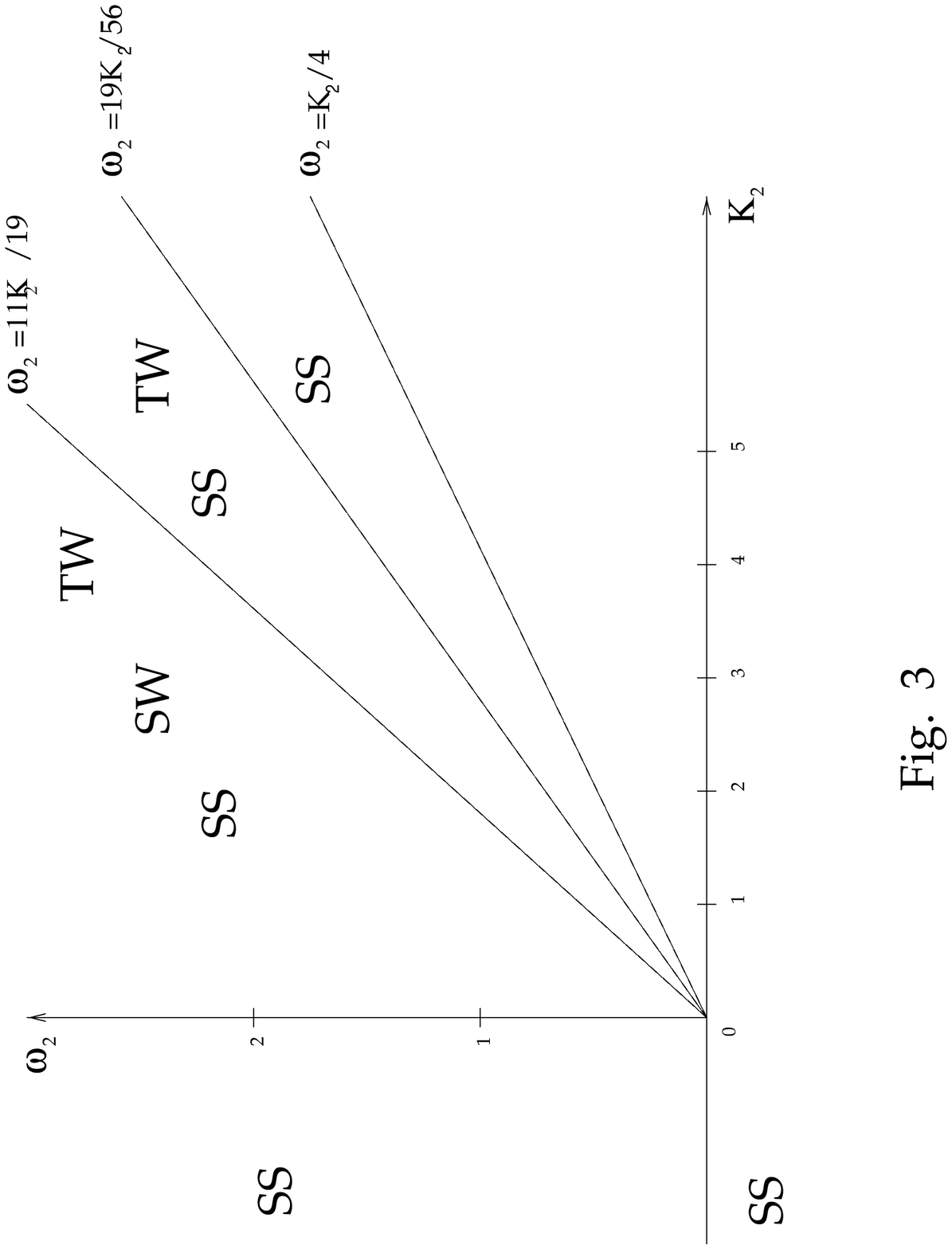,width=9.0cm,angle=-90}}}
\caption{Stability diagrams $(K_2,\omega_2)$ near the tricritical point.}
\end{figure}

\begin{figure}
\centerline{\hbox{\psfig{figure=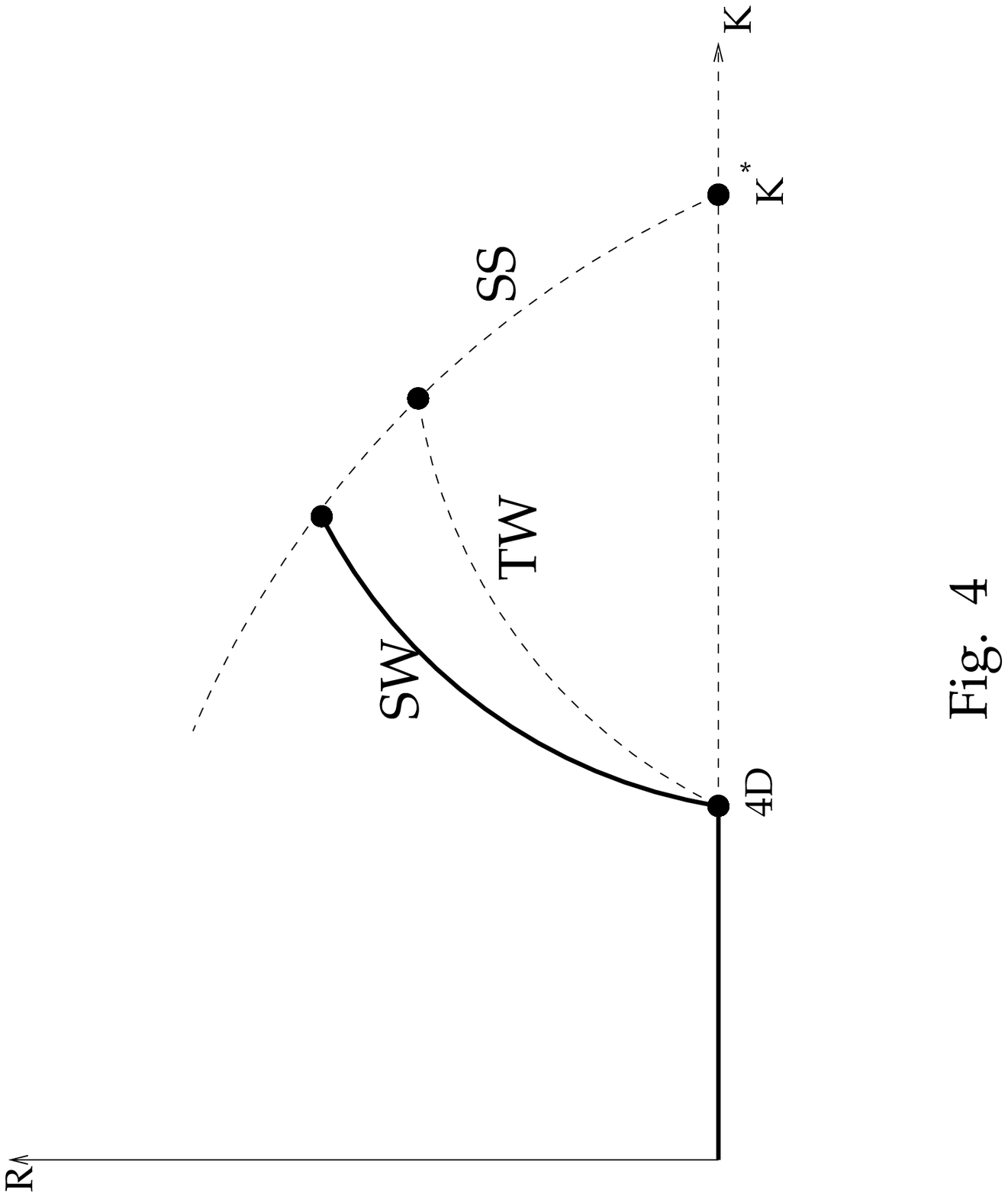,width=9.0cm,angle=-90}}}
\caption{Bifurcation diagram $(K,R)$ near the tricritical point for 
$\omega_0 > D$ fixed. $K^*$ is the coupling at which a subcritical
branch of stationary solutions bifurcates from incoherence.}
\end{figure}

\begin{figure}
\centerline{\hbox{\psfig{figure=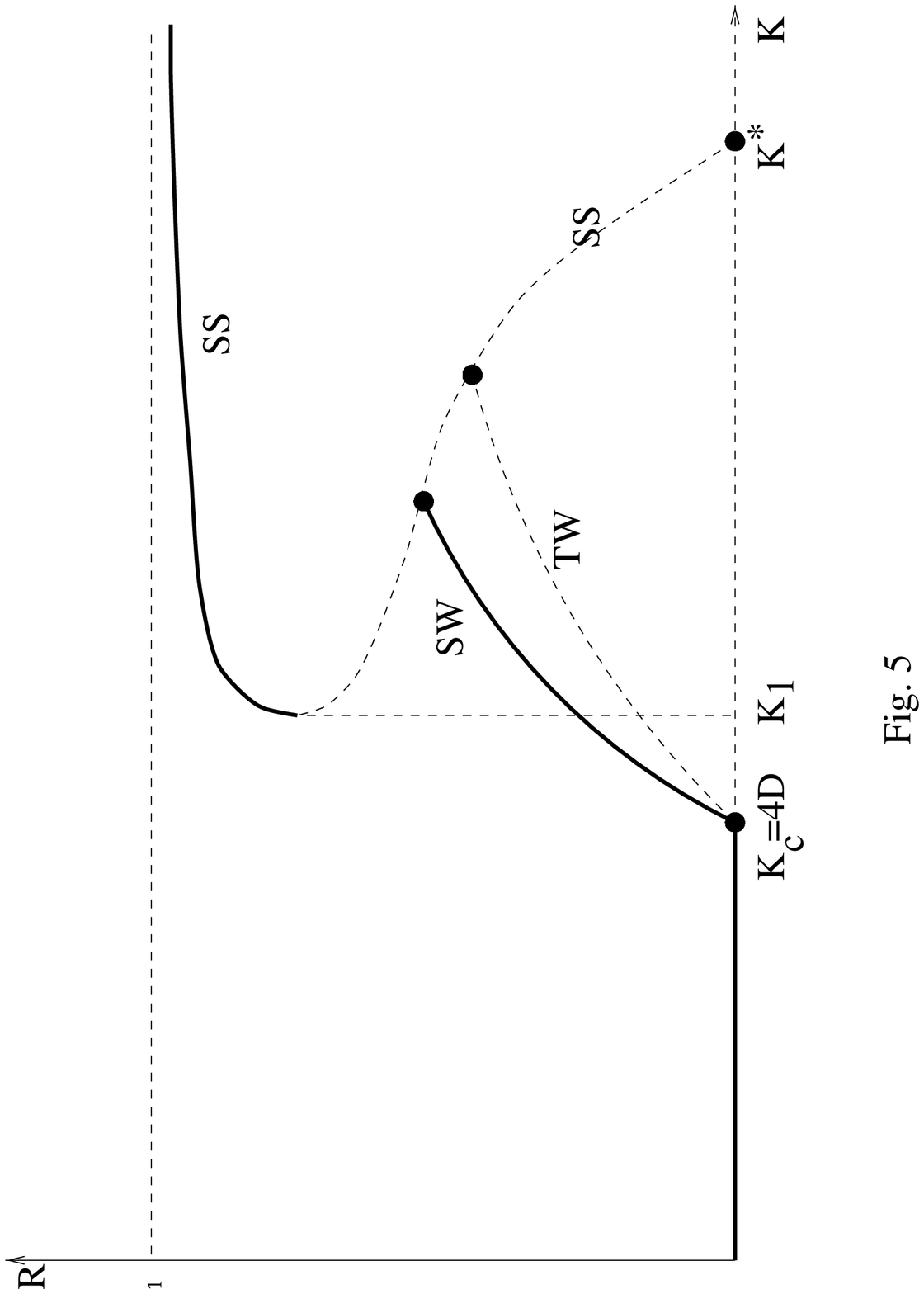,width=15.0cm,angle=-90}}}
\caption{Global bifurcation diagram including all stationary solution
branches for $\omega_0 > D$ fixed as conjectured from the information 
on bifurcating branches available near the tricritical point. The location 
of the turning point $K=K_1$ depends on the actual value of $\omega_0$.
The exchange of stabilities at the turning point is postulated, not 
demonstrated. Numerical simulations show that there is a narrow region 
of bistability between the SW and upper SS branches for $\omega_0 = 1.5 D$.
This region vanishes for $\omega_0 = 2D$.}
\end{figure}

\begin{figure}
\centerline{\hbox{\psfig{figure=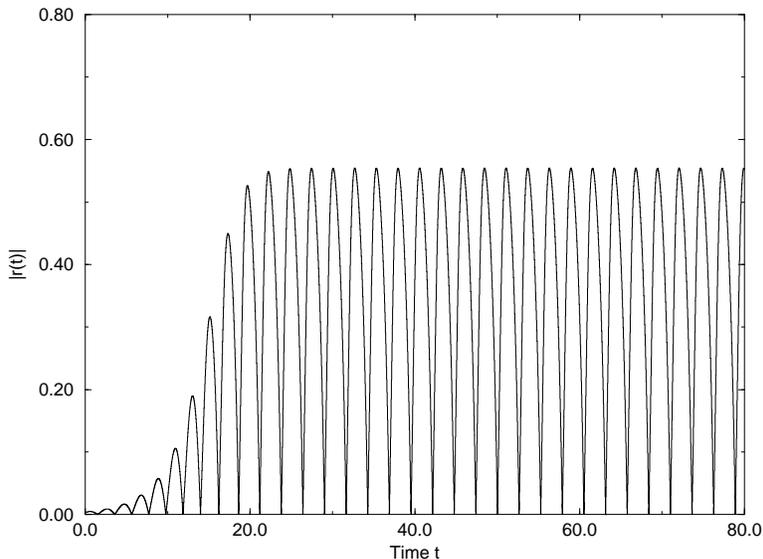,width=10.0cm,angle=-90}}}
\caption{Time evolution of the order parameter $|r(t)|$ for coupling strength
$K=5.2$, and $D=1$, $\omega_0=2$. Time is measured
in seconds, where one second means 200 time steps. We have considered as 
initial condition the incoherent solution $\rho\equiv 1/2 \pi$.}
\end{figure}

\begin{figure}
\centerline{\hbox{\psfig{figure=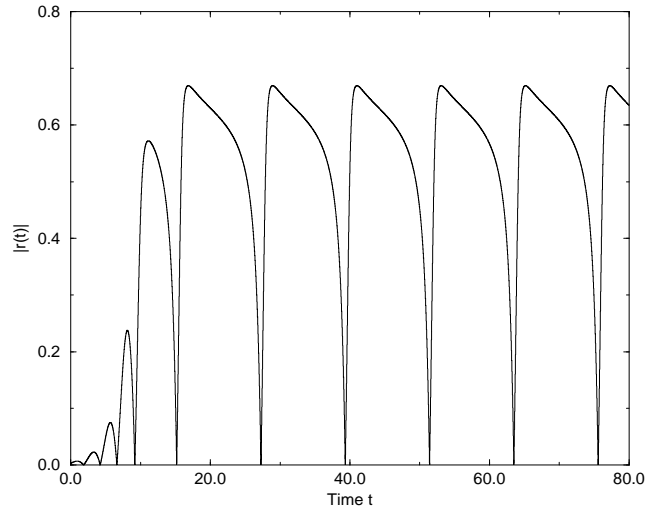,width=10.0cm,angle=-90}}}
\caption{Time evolution of the order parameter $|r(t)|$ for a larger value
of the coupling constant, $K=6$. As in the previous case there are oscillations 
but now their amplitude is larger as well as the period.}
\end{figure}

\begin{figure}
\centerline{\hbox{\psfig{figure=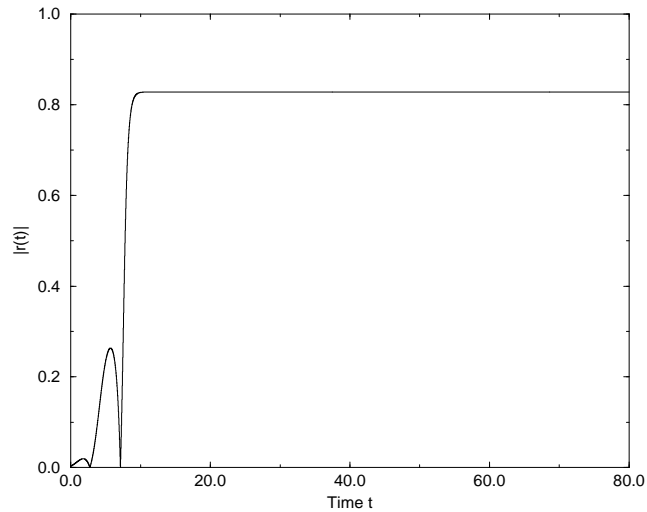,width=10.0cm,angle=-90}}}
\caption{Time evolution of the order parameter towards the stable synchronized
stationary solution for $\omega_0 = 2$, $D=1$ and $K=7$.}
\end{figure}

\begin{figure}
\centerline{\hbox{\psfig{figure=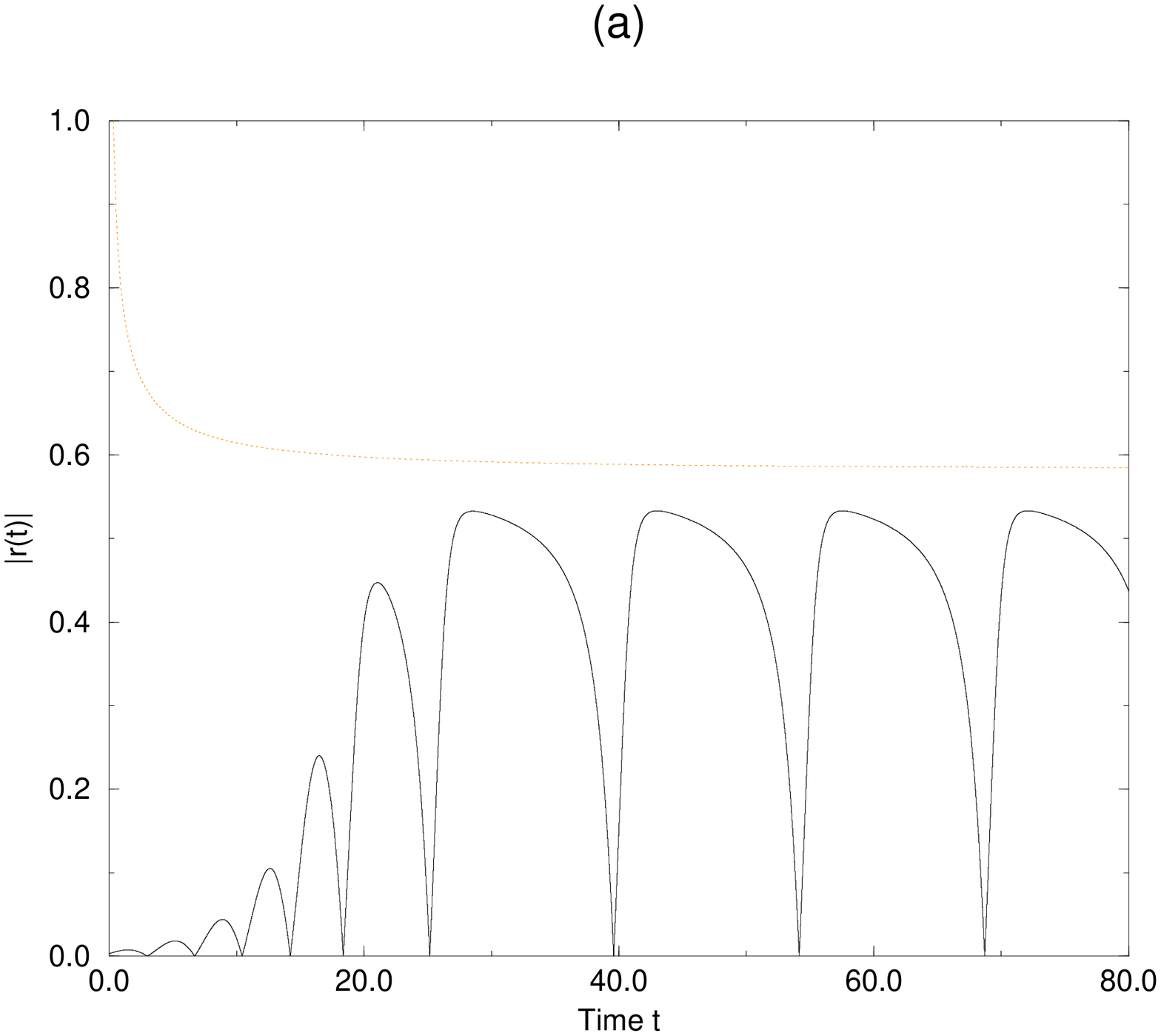,width=10.0cm,angle=-90}}}
\centerline{\hbox{\psfig{figure=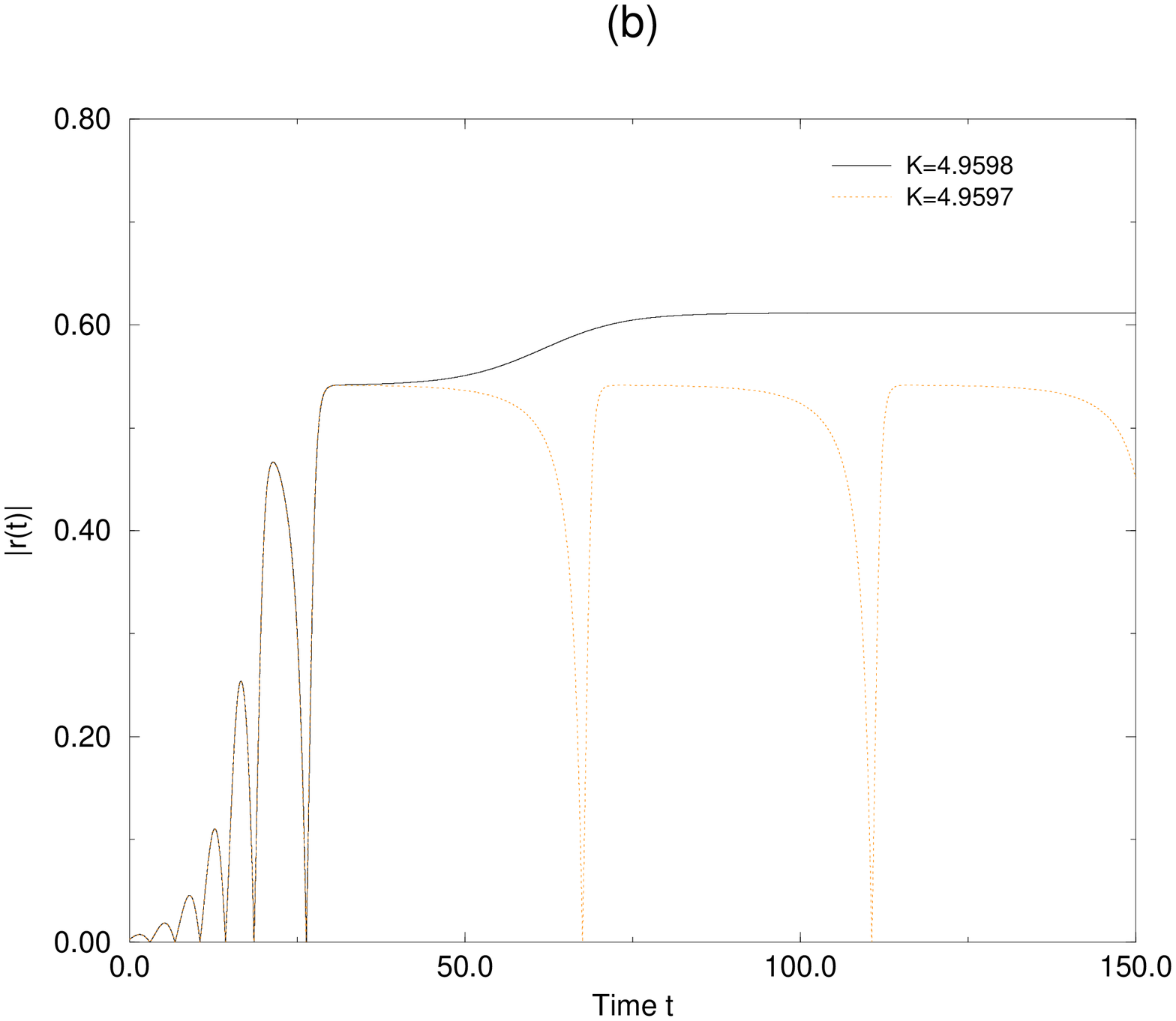,width=10.0cm,angle=-90}}}
\caption{(a) Time evolution of the order parameter in the parameter region
$\omega_0 = 1.5$, $D=1$ and $K=4.95$ where SW and SS solutions are both stable:
different initial data evolve to one of these solutions. (b) Details on the
end of the SW solution branch and abrupt transition to the SS at 
$K=4.9598$.}
\end{figure}

\begin{figure}
\centerline{\hbox{\psfig{figure=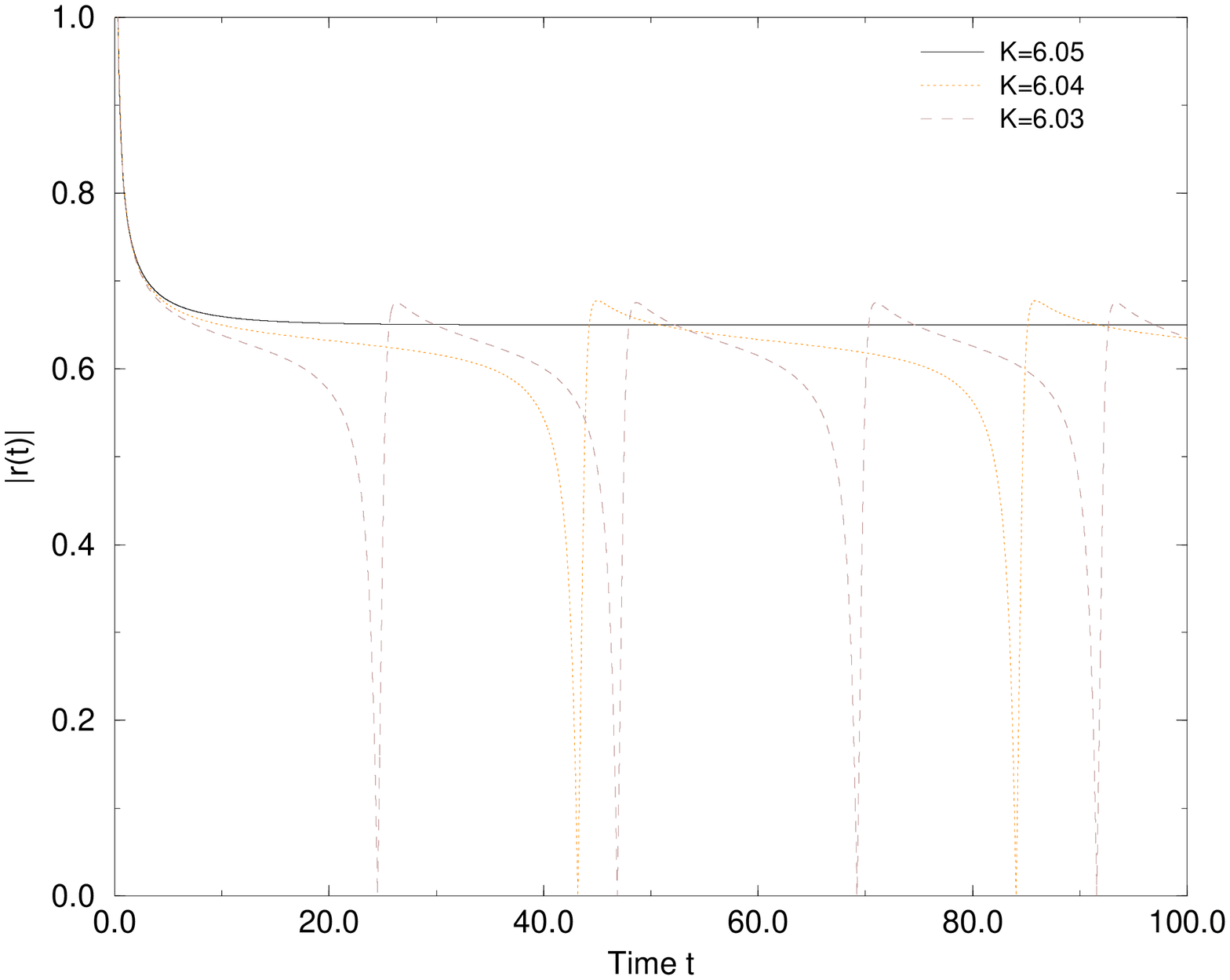,width=10.0cm,angle=-90}}}
\caption{Time evolution of the order parameter in the parameter region
$\omega_0 = 2$, $D=1$ for $K=6.03, 6.04, 6.05$ illustrating the smooth
transition between the stable SW and upper SS solution branches. For this 
value of $\omega_0$ there is no bistability between SW and SS solutions.}
\end{figure}


\begin{thebibliography}{99}
\bibitem[*]{bonilla:email} {E-address \tt bonilla@ing.uc3m.es}. Author to
whom all correspondence should be addressed.

\bibitem[\dagger]{cjpv:email} {E-address \tt conrad@ulyses.ffn.ub.es}

\bibitem[\ddag]{spigler:email} {E-address \tt spigler@ulam.dmsa.unipd.it}

\bibitem{TS1} K.Y. Tsang, S.H. Strogatz, and K. Wiesenfeld, {\it Reversibility
and noise sensitivity of Josephson arrays}, Phys. Rev. Lett. {\bf 66}, (1991),
1094-1097.

\bibitem{TS2} K.Y. Tsang, R.E. Mirollo, S.H. Strogatz, and K. Wiesenfeld, {\it
Dynamics of a globally coupled oscillator array}, Physica D {\bf 48}, (1991),
102-112.

\bibitem{CDW} S.H. Strogatz, C.M. Marcus, R.M. Westervelt and R.E.
Mirollo, {\it Simple model of collective transport with phase slippage}.
Phys. Rev. Lett. {\bf 61}, (1988) 2380-2383.

\bibitem{WIN} A.T. Winfree, {\it Biological rhythms and the behavior of
populations of coupled oscillators}, J. Theoret. Biol. {\bf 16}, (1967), 15-42.

\bibitem{STRO} S. H. Strogatz, {\it Norbert Wiener's brain waves}, edited
by S. Levin. Lect. N. Biomath. {\bf 100}, Springer, N. Y. 1994.

\bibitem{MISTRO} R. E. Mirollo and S. H. Strogatz, {\it Synchronization of the
pulse-coupled biological oscillators}, SIAM J. Appl. Math. {\bf 50}, (1990),
1645-1662.

\bibitem{GRAY} C.M. Gray and W. Singer, {\it Stimulus specific neuronal 
oscillations in the cat visual cortex: a cortical functional unit}, Soc. 
Neurosci. Abst. {\bf 13}, (1987) 13.

\bibitem{KURAM2} Y. Kuramoto, {\it Chemical Oscillations, Waves and
Turbulence}. Springer, N. Y. 1984.

\bibitem{ARO} D. G. Aronson, G. B. Ermentrout and N. J. Kopell, {\it Amplitude
response of coupled oscillators}, Physica D {\bf 41}, (1990) 403-449.

\bibitem{KURAM}  Y. Kuramoto, {\it Self-entrainment of a population of coupled
nonlinear oscillators}; in {\it International Symposium on Mathematical
Problems in Theoretical Physics}, H. Araki ed., Lecture Notes in Physics {\bf
39}, Springer, N. Y. 1975. pp.\ 420-422.

\bibitem{SAK} H. Sakaguchi, {\it Cooperative phenomena in coupled
oscillator systems under  external fields}. Prog. Theor. Phys. {\bf 79},
(1988) 39-46. 

\bibitem{STROMI2} S. H. Strogatz and R. E. Mirollo, {\it Phase-locking and 
critical phenomena in lattices of coupled nonlinear oscillators with random 
intrinsic frequencies}. Physica D {\bf 31}, (1988) 143-168.

\bibitem{LUMER} E. D. Lumer and B. A. Huberman, {\it Hierarchical dynamics in 
large assemblies of interacting oscillators}. Phys. Lett. A {\bf 160}, (1991)
 227-230.

\bibitem{BON2} L. L. Bonilla and J. M. Casado, {\it Dynamics of a soft-spin 
van Hemmen model. I. Phase and bifurcation diagrams for stationary 
distributions}. J. Statist. Phys. {\bf 56}, (1989) 113-125. 

\bibitem{PAB} C. J. P\'erez Vicente, A. Arenas and L.\ L.\ Bonilla, {\it
On the short time dynamics of networks of Hebbian coupled oscillators}. J.
Phys. A {\bf 29}, (1996) L9-L16.

\bibitem{BON3} L.L. Bonilla, C.J. P\'erez Vicente and J.M. Rub\'{\i}, {\it
Glassy synchronization in a population of coupled oscillators}. J. Stat.
Phys. {\bf 70}, (1993) 921-937.

\bibitem{SOMP} H. Sompolinsky, D. Golomb and D. Kleinfeld, {\it Cooperative 
dynamics in visual processing}. Phys. Rev. A
{\bf 43}, (1991) 6990-7011.

\bibitem{BON1} L. L. Bonilla, {\it Stable Probability Densities and Phase 
Transitions for Mean-Field Models in the Thermodynamic Limit}, 
J. Statist. Phys. {\bf 46}, (1987) 659-678.

\bibitem{STROMI} S.H. Strogatz and R.E. Mirollo, {\it Stability of incoherence
in a population of coupled oscillators}, J. Statist. Phys. {\bf 63}, (1991), 
613-635.

\bibitem{BNS} L. L. Bonilla, J. C. Neu, and R. Spigler, {\it Nonlinear
stability of incoherence and collective synchronization in a population of
coupled oscillators}, J. Statist. Phys. {\bf 67}, (1992), 313-330.

\bibitem{CRAW} J.D. Crawford, {\it Amplitude expansion for instabilities in
populations of globally-coupled oscillators}, J. Statist. Phys. {\bf 74},
(1994), 1047-1084.

\bibitem{OKUDA} K. Okuda and Y. Kuramoto, {\it Mutual entrainment between
populations of coupled oscillators}, Prog. Theor. Phys. {\bf 86},
(1991), 1159-1176.

\bibitem{DK} G. Dangelmayr and E. Knobloch, {\it The Takens-Bogdanov
bifurcation with the O(2)-symmetry}, Phil. Trans. R. Soc. Lond. A {\bf 322},
(1987), 243-279.

\bibitem{LIN} 
J. Lin and P. B. Kahn, {\it Averaging methods in the delayed logistic equation}. 
J. Math. Biol. {\bf 10} (1980), 89-96.
 J. Lin and P. B. Kahn, {\it Phase and amplitude instability in delay-diffusion 
population models}. J. Math. Biol. {\bf 13} (1982), 383-393.

\bibitem{Cole} J.~D.\ Cole and J.\ Kevorkian, {\it Multiple Scale and Singular 
Perturbation Methods}, Springer, New York, 1996.

\end{thebibliography}
\end{document}